\documentclass[twocolumn,superscriptaddress,prb,floatfix,twoside,showpacs,bibnotes]{revtex4-1}
\pdfoutput=1
\usepackage{amsmath,amsfonts,graphics,multirow}

\IfFileExists{microtype.sty}{\usepackage{microtype}}{\relax}

\usepackage{hyperref}
\hypersetup{
 pdfpagemode=UseNone,
 pdfstartview=FitH
}

\def\rmd{{\rm d}}
\def\rmi{{\rm i}}
\def\rme{{\rm e}}
\def\im{\mathop{\rm Im}\nolimits}

\def\tr{\mathop{\rm Tr}\nolimits}

\def\selfen{\Sigma}
\def\bbselfen{\Sigma}

\def\code#1{\textsc{\MakeLowercase{#1}}}

\begin{document}

\title{Electronic structure and core-level spectra of light actinide
  dioxides in the dynamical mean-field theory}

\author{Jind\v{r}ich Koloren\v{c}}\email{kolorenc@fzu.cz}
\affiliation{Institute of Physics, Academy of Sciences of
the Czech Republic, Na Slovance 2, CZ-182 21 Praha 8, Czech Republic}

\author{Alexander B. Shick}
\affiliation{Institute of Physics, Academy of Sciences of
the Czech Republic, Na Slovance 2, CZ-182 21 Praha 8, Czech Republic}

\author{Alexander I. Lichtenstein}
\affiliation{Institut f\"ur Theoretische Physik, Universit\"at
  Hamburg, Jungiusstra\ss e 9, D-20355 Hamburg, Germany}

\date{\today}

\begin{abstract}
The local-density approximation combined with the dynamical mean-field
theory (LDA+DMFT) is applied to the paramagnetic phase of light actinide
dioxides: UO${}_2$, NpO${}_2$, and PuO${}_2$. The calculated band gaps
and the valence-band electronic structure are in a very good agreement
with the optical absorption experiments as well as with the
photoemission spectra. The hybridization of the actinide 5f shell with
the 2p states of oxygen is found to be relatively large, it increases
the filling of the 5f orbitals from the nominal
ionic configurations with two, three, and four electrons to nearly
half-integer values 2.5, 3.4 and 4.4. The large hybridization leaves
an imprint also on the core-level photoemission spectra in the form
of satellite peaks. It is demonstrated that these satellites are
accurately reproduced by the LDA+DMFT calculations.
\end{abstract}


\maketitle


\section{Introduction}

Actinide dioxides are correlations-driven
insulators~\cite{schoenes1978,mccleskey2013} that display a variety of
complex ordered phases at low temperatures.\cite{santini2009} They
crystallize in the CaF$_2$ structure (space group Fm$\bar{3}$m), with
eight-coordinated actinide atoms, and four-coordinated oxygen
atoms. Due to the interplay of electron correlations, spin-orbital
coupling and crystal field effects, the theoretical modeling of the
electronic structure of these oxides presents numerous challenges.

The conventional Kohn--Sham density-functional theory (DFT) in the
local-density (LDA) and generalized gradient approximations fails to
explain the insulating character of the oxides.\cite{wen2013} The
band-gap problem was addressed a number of times using
orbital-dependent functionals such as the self-interaction corrected
LDA,\cite{petit2010} LDA+U,\cite{suzuki2013} or the hybrid
exchange-correlation functionals.\cite{prodan2007,wen2012,wen2013} All
of these calculations lead to insulators but the opening of the gap is
intimately linked with the appearance of a long-range magnetic
order. That is not satisfactory since the oxides retain the gap also
in the high-temperature paramagnetic phase. Moreover, the
orbital-dependent functionals predict a large magnetic moment at the
plutonium atoms in PuO${}_2$ in disagreement with
experiments.\cite{raphael1968} These issues appear to be a rather
general shortcoming of static mean-field approximations that build on
a single determinant of Kohn--Sham orbitals.

The dynamical mean-field theory (DMFT) is able to describe correlated
nonmagnetic insulators.\cite{georges1996} This method, in combination
with the density-functional theory, was successfully applied to
selected actinide dioxides recently,\cite{yin2011,shick2014} and it
indeed yields an insulating electronic structure without any
long-range order and without any local magnetic moments in PuO${}_2$. In
the present paper, we follow up on our previous study of the plutonium
dioxide,\cite{shick2014} where we employed a crystal-field potential
deduced from experiments and assumed a simplified spherically
symmetric hybridization of the plutonium 5f shell with the surrounding
electronic states. Here we relax these simplifications and determine
the quantities entirely from first principles. We investigate also the
paramagnetic phases of UO${}_2$ and NpO${}_2$ in order to
visualize trends in the behavior of the computed properties when the
filling of the actinide 5f shell changes.

\section{Methods}

\subsection{LDA}

\begin{figure}
\resizebox{0.8\linewidth}{!}{\includegraphics{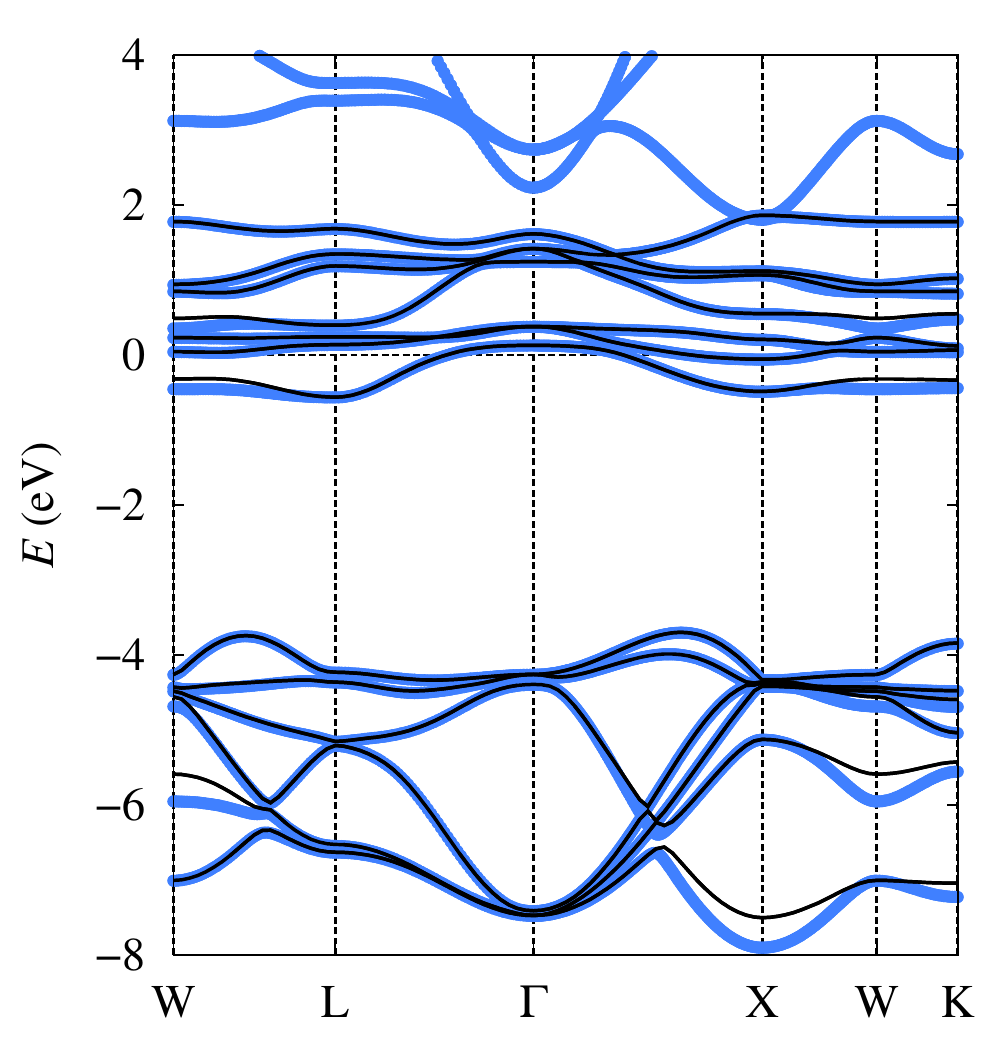}}
\caption{\label{fig:UO2_LDA_bands}(Color online) LDA band structure of
  UO${}_2$ from \code{WIEN2k} (thick blue line) and after mapping onto
  the tight-binding model $\hat H_{\rm TB}$
  (thin black line). The dominant orbital character is oxygen 2p
  between $-8$~eV and $-4$~eV, and uranium 5f between $-1$~eV and 2~eV. The
  visible bands above 2~eV are mostly uranium 6d.}
\end{figure}

We start with all-electron LDA calculations\cite{perdew1992}
of the band structure of the dioxides at the experimental lattice
constants (Tab.~\ref{tab:nf_gap}). We take into account
scalar-relativistic effects as well as the spin-orbital coupling. To
this end, we employ the \code{WIEN2k} package\cite{wien2k} with the
following parameters: the radii of the muffin-tin spheres are
$R_{\rm MT}({\rm U})=R_{\rm MT}({\rm Np})=R_{\rm MT}({\rm Pu})
=2.65\, a_{\rm B}$ for actinide atoms and $R_{\rm
MT}({\rm O})=1.70\, a_{\rm B}$ for oxygen atoms, the basis-set cutoff
$K_{\rm max}$ is defined with $R_{\rm MT}({\rm O})\times K_{\rm
    max}=8.5$, and the
Brillouin zone is sampled with 3375 k points (120 k points in the
irreducible wedge). The computed electronic structure is essentially
identical to the nonmagnetic LDA results shown in
Ref.~\onlinecite{suzuki2013}.

The LDA bands of the actinide 5f and oxygen 2p 
origin are subsequently mapped onto a tight-binding model with the aid
of the \code{Wannier90} code\cite{mostofi2008} in conjunction with
the \code{Wien2wannier} interface.\cite{kunes2010} This effective
model $\hat H_{\rm TB}$ then serves as a base for the LDA+DMFT
calculations. The tight-binding representation is not perfect, there
are some deviations from the original band structure that
originate mainly in the neglected overlap of the actinide 6d states
with the oxygen 2p orbitals. The largest differences appear in
UO${}_2$, they are explicitly illustrated in
Fig.~\ref{fig:UO2_LDA_bands}. It is possible to get a closer
match even for UO${}_2$, but the resulting 5f Wannier functions have
a varied spatial extension which complicates their interpretation
as an atomic f~shell later on.

\subsection{LDA+DMFT}
\label{sec:DMFT}

The dynamical mean-field modeling of correlations among the 5f
electrons amounts to adding a local selfenergy $\hat\selfen(z)$ to the
5f shell of each actinide atom in the tight-binding model~$\hat H_{\rm
  TB}$. The selfenergy is taken from an
auxiliary impurity model that consists of one fully interacting f
shell (the impurity) embedded in a self-consistent non-interacting
medium $\bigl(\hat H_{\rm TB}+\hat\selfen\bigr)$.\cite{georges1996}

The auxiliary model without the f--f interactions can be written as
\begin{multline}
\label{eq:Himp0}
\hat H_{\rm imp}^{(0)}=
\sum_{ij}\bigl[\mathbb{H}_{\rm loc}\bigr]_{ij}
  \hat f_i^{\dagger}\hat f_j
+\sum_{IJ}\bigl[\mathbb{H}_{\rm bath}\bigr]_{IJ}
  \hat b_I^{\dagger}\hat b_J\\
+\sum_{iJ}\Bigl(\bigl[\mathbb{V}\bigr]_{iJ}\hat f_i^{\dagger}\hat b_J
+\bigl[\mathbb{V}^{\dagger}\bigr]_{Ji}\hat b_J^{\dagger}\hat f_i
\Bigr)\,,
\end{multline}
where the lower-case indices label the f orbitals and run from 1 to
14 (or they can be understood as combinations of the magnetic
quantum number~$m$ and the spin projection~$\sigma$), and the
upper-case indices label the orbitals of the effective
medium that is usually referred to as bath. We truncate the bath to
contain only 14 orbitals so that
the local hamiltonian~$\mathbb{H}_{\rm loc}$, the bath
hamiltonian $\mathbb{H}_{\rm bath}$ as well as the hybridization
$\mathbb{V}$ are all $14\times 14$ matrices. The actual determination
of $\mathbb{H}_{\rm loc}$, $\mathbb{H}_{\rm bath}$ and $\mathbb{V}$ is
discussed in detail in appendix~\ref{app:bath_discretization}. Here we
just note that they cannot be reduced to a diagonal form due to
non-commutativity of the cubic hybridization with the spin-orbital
coupling.

The truncation of the bath is necessary because the Lanczos method,
which we employ to solve the impurity model, cannot handle
much larger systems. In insulating oxides, the small bath is well
justified on the physical grounds: the environment of the 5f shell is
dominated by the oxygen ligands and hence a small impurity model
analogous to the ligand-field model should accurately represent the
dynamics of the 5f shell and its surroundings. High accuracy of these
small models was demonstrated many times in the context of core-level
spectroscopies,\cite{degroot2008} recently for instance in
Ref.~\onlinecite{haverkort2012}, as well as in applications to the
valence-band electronic structure of transition-metal
oxides.\cite{thunstrom2012}

The complete interacting impurity model is defined as $\hat
H_{\rm imp}=\hat H_{\rm imp}^{(0)}+\hat U$ where $\hat U$ is the
Coulomb repulsion among the f orbitals,
\begin{equation}
\label{eq:vertex}
\hat U=
\frac12\sum_{ijkl} U_{ijkl}
\hat f_{i}^{\dagger}\hat f_{j}^{\dagger}
\hat f_{l}\hat f_{k}\\
-\sum_{ij}\bigl[\mathbb{U}_{\rm H}\bigr]_{ij}\hat f_i^{\dagger}\hat f_j\,.
\end{equation}
The matrix elements $U_{ijkl}$ are decomposed into the radial
expectation values -- the Slater integrals $F_k$, and the angular
expectation values -- the Gaunt coefficients. The latter are fully
determined by the assumption that the angular parts of the local
orbitals are spherical harmonics. The Slater integrals $F_k$ are
assumed identical for all three oxides. They are set to $F_0=6.5$~eV,
$F_2=8.1$~eV, $F_4=5.4$~eV and $F_6=4.0$~eV. The average Coulomb
repulsion $U=F_0$ has a value close to Ref.~\onlinecite{yin2011} to
facilitate comparison with this earlier study. The remaining Slater
integrals are chosen to obtain the average exchange $J=0.7$~eV while
keeping the ratios $F_4/F_2$ and $F_6/F_2$ equal to their
Hartree--Fock values.\cite{moore2009}

 The second term in Eq.~\eqref{eq:vertex} is the double-counting
correction that removes the Hartree-like contribution already included
in the LDA band structure $\hat H_{\rm TB}$. In the paramagnetic phase,
the correction can be simplified to an isotropic form,
$\bigl[\mathbb{U}_{\rm H}\bigr]_{ij}=U_{\rm H}\delta_{ij}$. We
approximate $U_{\rm H}$ by the so-called fully localized limit $U_{\rm
H}=U(n_f-1/2)-J(n_f-1)/2$, where $n_f$ is the self-consistently
determined number of 5f electrons.\cite{czyzyk1994,solovyev1994} The
isotropic form of the double counting is still only approximate even
in the paramagnetic state. It is accurate enough for getting the
correct position of the 5f states with respect to the ligand bands,
but it is possibly insufficient when it comes to crystal-field effects
that occur at a much smaller energy scale, especially in the case of
4f electrons.\cite{novak2013}

The selfenergy $\hat\selfen(z)$ in the impurity model $\hat H_{\rm
imp}$ is computed using an in-house exact-diagonalization code that
combines the implicitly restarted Lanczos method for calculation of
the bottom of the many-body spectrum \cite{arpack} with the band
Lanczos method for evaluation of the one-particle Green's
function.\cite{meyer1989} The calculations are performed at room
temperature ($T=300$~K), well within the paramagnetic phase. To lessen
the computational demands, the Fock space is reduced in a manner
analogous to the method developed for Ce compounds by Gunnarsson and
Sch\"onhammer.\cite{gunnarsson1983} Details of the impurity solver are
discussed in appendix~\ref{app:solver}.

\subsection{Photoemission spectra}
\label{sec:spectra_theo}

The valence-band photoemission intensity can be evaluated
using the Fermi's golden rule. If the energy and angular dependence of
the dipole matrix elements is neglected, the angle-resolved
photoelectron spectrum $I_{\rm v}(k,E)$ is proportional to the
one-particle spectral density $A(k,E)$ 
of the tight-binding model,\footnote{We
  use the following simplified notation: wherever we
  add a scalar to an operator or to a matrix, it is to be understood
  as adding the scalar only to the diagonal elements. That is, $z+\hat H\equiv
  z\hat I+\hat H$ and $z+\mathbb{H}\equiv z\mathbb{I}+\mathbb{H}$,
  where $\hat I$ and $\mathbb{I}$ stand for the identity operator and the
  identity matrix.}
\begin{equation}
\label{eq:valence_spectral_dens}
A=\frac1{\pi}\im\tr\biggl[
\frac1{%
E-\rmi \Gamma_{\rm v}-\hat H_{\rm TB}(k)
-\hat\selfen(E-\rmi \Gamma_{\rm v})}\biggr]\,.
\end{equation}
A finite imaginary part of the one-particle energy, $\Gamma_{\rm v}$, was
introduced to model the life-time broadening of the valence
states together with the experimental resolution. The trace in
Eq.~\eqref{eq:valence_spectral_dens} runs over
all valence states or over a subset of orbitals (actinide 5f or oxygen 2p)
if an orbital-resolved signal is desired. The
angle-integrated photoelectron spectrum $I_{\rm v}(E)$ is proportional
to the integral of the spectral density $A(k,E)$ over the first
Brillouin zone.

Apart from the valence electronic structure, the impurity model of
the dynamical mean-field theory provides a means to calculate also the
photoemission from core states (x-ray photoemission spectra,
XPS)\cite{kim2004,cornaglia2007,hariki2013} at the level of the
so-called multiplet ligand-field theory.\cite{degroot2008} To that
end, we add a core state $\hat c$ and the free-electron states $\hat
a_k$ to the impurity model,
\begin{equation}
\label{eq:HimpXPS}
\hat H_{\rm XPS}=\hat H_{\rm imp}+\epsilon_{\rm c}\hat c^{\dagger}\hat c
+U_{\rm cv}(\hat c^{\dagger}\hat c-1)\hat n_f
+\sum_k\epsilon_k\hat a_k^{\dagger}\hat a_k\,,
\end{equation}
where $\epsilon_{\rm c}$ is the energy of the core state, $\epsilon_k$
are the energies of the free-electron states, $\hat n_f=\sum_i\hat
f_i^{\dagger}\hat f_i$ is the number of f electrons, and $U_{\rm cv}$ is the
strength of the Coulomb repulsion between the core electrons and the f
electrons. The degeneracy of the core state is neglected and hence
there is only one core-valence Slater integral.

Employing the Fermi's golden rule, the probability $I_{\rm
  c}(E)$ of the emission of an electron from the core level to a
scattering state with an energy $E$ can be written
as\cite{gunnarsson1983,degroot2008}
\begin{equation}
\label{eq:core_spectrum}
I_{\rm c}\sim\im\,
\langle0|
\frac1{%
E-\rmi\Gamma_{\rm c}-E_0-\epsilon_{\rm c}
+\hat H_{\rm imp}-U_{\rm cv}\hat n_f
}
|0\rangle\,,
\end{equation}
where $|0\rangle$ is the many-body ground state of the converged DMFT
impurity model $\hat H_{\rm imp}$ (that is, the initial state of the
photoemission process), $E_0$ is the corresponding ground-state
energy, and $\Gamma_{\rm c}$ simulates the life-time broadening of the
core state. As in the case of the valence spectra, we neglected the
energy dependence of the dipole matrix elements. In addition, we
assumed that the free-electron density of states is a constant in the
energy window of interest.

The expression shown in Eq.~\eqref{eq:core_spectrum} applies at zero
temperature, $T=0$~K. The finite-temperature case is analogous to the
finite-temperature formula for the one-particle Green's function,
Eqs.~\eqref{eq:GFgrandcanonical}, discussed in
appendix~\ref{app:solver}. Just like the Green's function, the
photoemission intensity $I_{\rm c}(E)$ is computed using the Lanczos
method.

In the course of derivation of Eq.~\eqref{eq:core_spectrum},
the $\hat c$ and $\hat a_k$ degrees of freedom were integrated out and
hence the final-state hamiltonian $\hat H_{\rm final}=\bigl(\hat H_{\rm
  imp}-U_{\rm cv}\hat n_f\bigr)$, which enters the denominator of
Eq.~\eqref{eq:core_spectrum}, involves only the f
orbitals and the ligand states. Notably, $\hat H_{\rm final}$ has the
same form as $\hat H_{\rm imp}$, only the diagonal elements of
$\mathbb{H}_{\rm loc}$ are altered. To make the model more realistic,
we scale down the hybridization parameters in the final-state
hamiltonian $\hat H_{\rm final}$, $\mathbb{V}\to q\mathbb{V}$,
$0<q<1$. The hybridization between the f shell and the ligand orbitals
is reduced due to a contraction of the 5f wave functions when the core hole
is present.\cite{gunnarsson1988,degroot2008} The reduction of the
hybridization parameters can be incorporated into the hamiltonian of
Eq.~\eqref{eq:HimpXPS} by introducing an extra term
\begin{equation}
(q-1)\,(\hat c^{\dagger}\hat c-1)
\sum_{iJ}\Bigl(\bigl[\mathbb{V}\bigr]_{iJ}\hat f_i^{\dagger}\hat b_J
+\bigl[\mathbb{V}^{\dagger}\bigr]_{Ji}\hat b_J^{\dagger}\hat f_i
\Bigr)\,.
\end{equation}
This term does not have any transparent physical interpretation but it can
still be used to formally check that the rescaling $\mathbb{V}\to
q\mathbb{V}$ in the final state is compatible with
the steps that led from Eq.~\eqref{eq:HimpXPS} to
Eq.~\eqref{eq:core_spectrum}.


\section{Results and discussion}


\begin{table}
\caption{\label{tab:nf_gap}%
Basic characteristics of the investigated dioxides: the experimental lattice
constant $a$ (Ref.~\onlinecite{PearsonsHandbook}), the number of 5f
electrons $n_f$ calculated by a projection onto the Wannier
orbitals, and the computed band gaps compared to the experimental
values.}
\begin{ruledtabular}
\begin{tabular}{lccccc}
& \multirow{2}{*}{$a$ (\AA)} & \multicolumn{2}{c}{$n_f$}
& \multicolumn{2}{c}{gap (eV)} \\
& & LDA & DMFT & theory & experiment  \\
\hline
UO${}_2$  & 5.470 & 2.8 & 2.5 & 1.9 & 2.1 (Ref.~\onlinecite{schoenes1978})\\
NpO${}_2$ & 5.432 & 3.8 & 3.4 & 2.5 & 2.8 (Ref.~\onlinecite{mccleskey2013})\\
PuO${}_2$ & 5.396 & 4.8 & 4.4 & 2.5 & 2.8 (Ref.~\onlinecite{mccleskey2013})\\
\end{tabular}
\end{ruledtabular}
\end{table}

\subsection{Valence-band electronic structure}

\begin{figure*}[t]
\resizebox{.9\linewidth}{!}{\includegraphics{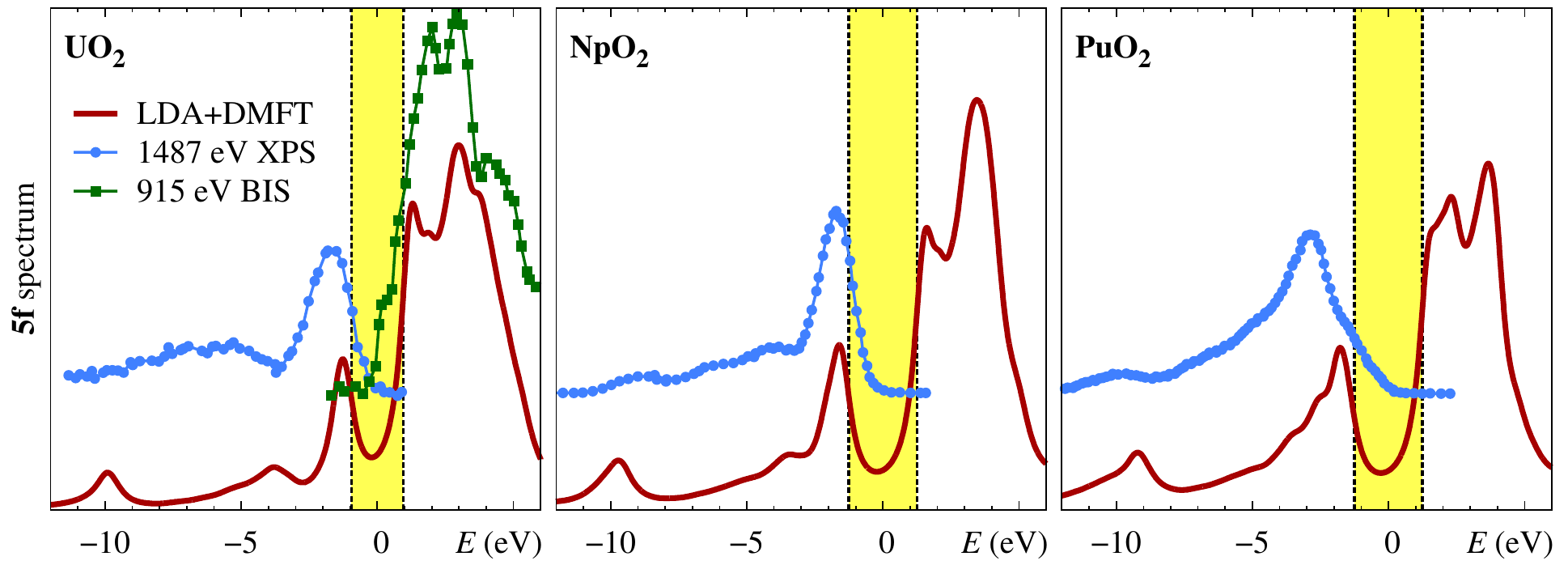}}
\caption{\label{fig:5fdos}(Color online) The $k$-integrated spectral density
  corresponding to the 5f states (life-time broadening
  $\Gamma_{\rm v}=0.4$~eV, thick red line) is
  compared  with the x-ray photoemission spectra at the 
  aluminum $K_{\alpha}$ line (connected blue dots). The experimental
  data are adopted from Ref.~\onlinecite{yu2011} (UO${}_2$),
  Ref.~\onlinecite{teterin2014} (NpO${}_2$), and Ref.~\onlinecite{teterin2013}
  (PuO${}_2$). The yellow stripes indicate where the band gap would appear
  if the broadening were removed. The Fermi level is placed in the
  center of the gap.}
\end{figure*}

\begin{figure}
\resizebox{.93\linewidth}{!}{\includegraphics{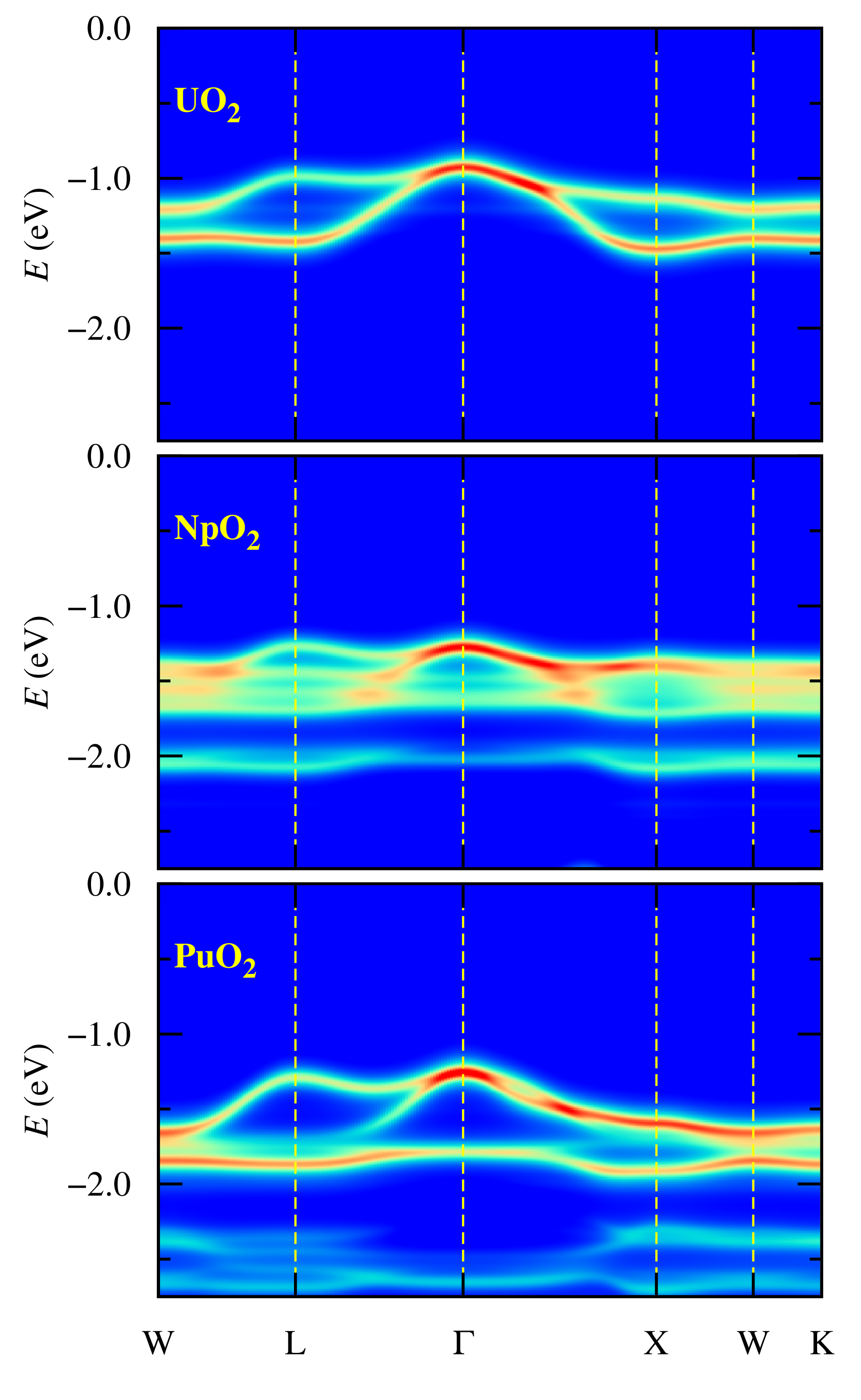}}
\caption{\label{fig:5fbands}(Color online) The spectral density 
  of the 5f states in the first $2.75$~eV below the Fermi
  level. The life-time broadening is smaller than in Fig.~\ref{fig:5fdos}
  ($\Gamma_{\rm v}=0.08$~eV) in order to resolve finer details of the
  band structure.}
\end{figure}

The LDA+DMFT calculations result in an insulating electronic structure
for all three investigated dioxides. The obtained band gaps are listed
in Tab.~\ref{tab:nf_gap} together with the corresponding experimental
data. The calculations slightly underestimate the gaps, which could be
straightforwardly fixed by an increased value of the Coulomb
$U$. Nevertheless, we get the correct trend in the magnitude of the
gap across the three oxides even without any such fine tuning, simply
with the same Coulomb parameters in all cases. This indicates that the
material-specific screening of the Coulomb interaction is accurately
represented already in the employed minimal model of the valence
electronic structure. An analogous conclusion was reached in a recent
LDA+DMFT study of transition-metal oxides.\cite{haule2014}

The $k$-integrated spectral density corresponding to the actinide 5f
states is shown in Fig.~\ref{fig:5fdos} in comparison with the
angle-integrated x-ray photoelectron spectra measured with incident
photons at the aluminum $K_{\alpha}$ line (1487~eV). At this energy,
the photoionization cross section for the oxygen 2p states is
negligible and hence the experimental data contain essentially
clean 5f signal. The total spectral density is measured with incident
photons at He~II line (40.8~eV) as their cross section with actinide
5f and oxygen 2p states is approximately equal. Comparison of He~II spectra
of PuO${}_2$ with our LDA+DMFT calculations is shown in
Ref.~\onlinecite{kolorenc2015}.

Inspecting the case of UO${}_2$ in Fig.~\ref{fig:5fdos} first, we see
three distinct features
in the occupied part of the theoretical spectrum: the main 5f bands
near $-1$~eV, a satellite peak near $-10$~eV, and a broader feature
between $-3$~eV and $-7$~eV. The last feature coincides with the location
of the oxygen 2p bands and reflects the hybridization of the 5f states
with the ligands. The three features are discernible also in the spectra
of NpO${}_2$ and PuO${}_2$, only the 2p bands move closer to the main
5f peak, and eventually overlap with this peak in PuO${}_2$. The
theoretical calculations closely reproduce the shape of the
experimental spectra as well as their evolution from UO${}_2$ to
PuO${}_2$, which demonstrates the accuracy of the LDA+DMFT modeling of the
electronic structure.

The momentum-resolved spectral density is shown as a color map in
Fig.~\ref{fig:5fbands}. The dispersion of the 5f bands is approximately
0.5~eV in UO${}_2$ and it increases only a little to about 0.75~eV in
PuO${}_2$. The hybrid DFT calculations predict a much larger
increase of the 5f bands dispersion in PuO${}_2$ due to the degeneracy of
the plutonium 5f and oxygen 2p bands (Fig.~\ref{fig:5fdos}) and a
consequent enhancement of the effects of
hybridization.\cite{prodan2007,wen2012,wen2013} Our 
calculations suggest that such increase of the 5f states dispersion is an
artifact of the single-determinant approximation employed in the DFT
calculations. Although the angular-resolved photoemission (ARPES)
experiments were performed on PuO${}_2$, the acquired data are not
able to resolve the issue yet due to the lack of orbital resolution
(the measurements were done with the He~II light
source).\cite{joyce2010} The analysis of ARPES data is more
straightforward in UO${}_2$ where the 5f bands are well separated and
it is possible to read out their dispersion.\cite{roy2008} It comes
out as about 0.13~eV, which is considerably smaller than our result
plotted in Fig.~\ref{fig:5fbands}. A large part of the
discrepancy can be attributed to the smearing of the experimental data
that merges the two bands into a single peak with an apparent
dispersion approximately 50\% smaller than the resolved bands. The
multiband composition of this single peak shows up in the
experiment as a variation of the peak's shape along the momentum path.

\subsection{Screening of the Coulomb interaction}

\begin{figure}
\resizebox{.9\linewidth}{!}{\includegraphics{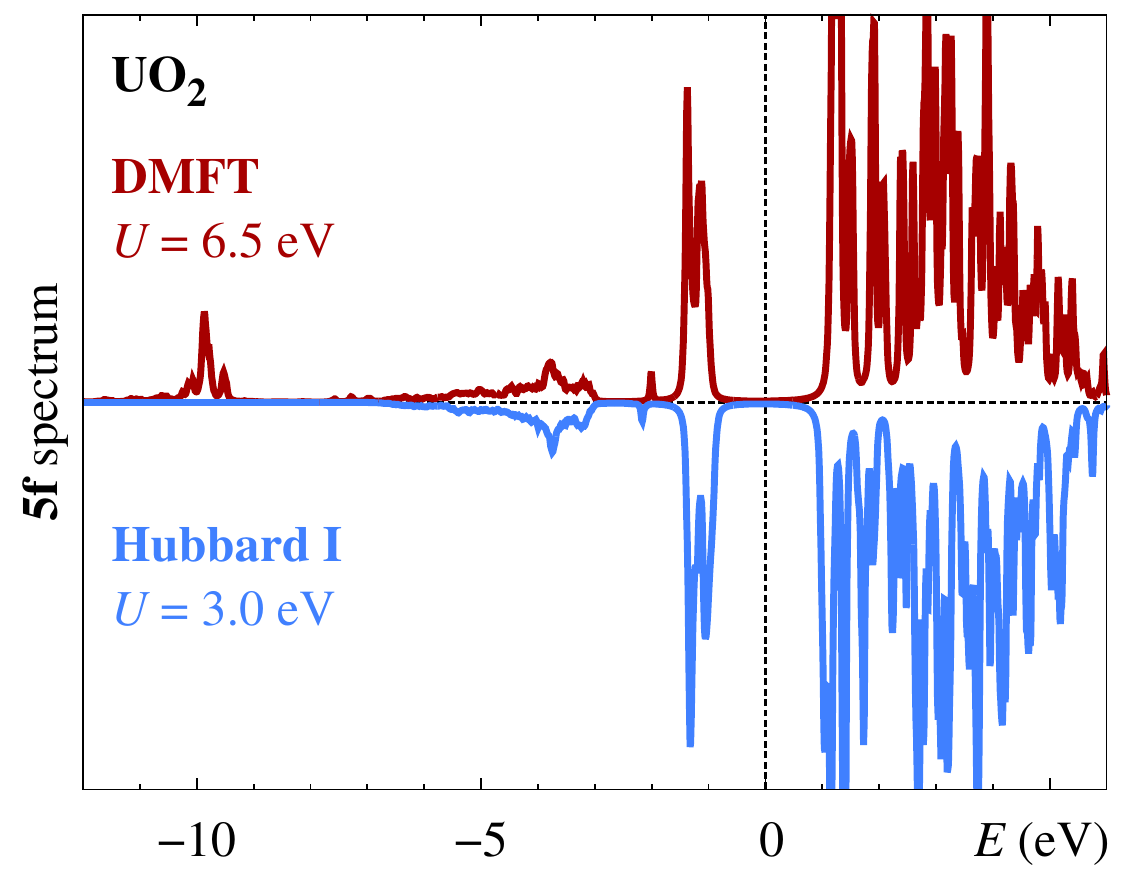}}
\caption{\label{fig:UO2_dmft_hub1}(Color online) The spectral density of the 5f
  states calculated with the DMFT selfenergy and $U=6.5$~eV (red,
  top), and with the selfenergy from the Hubbard~I approximation and
  $U$ reduced to $3.0$~eV (blue, bottom). The
 only larger difference is the absence of the $-10$~eV satellite in
 the  Hubbard~I calculations.}
\end{figure}

It was demonstrated in the previous section and also in
Ref.~\onlinecite{yin2011} that the LDA+DMFT method yields an accurate
electronic structure if the Coulomb $U$ is set around $6.5$~eV, or
perhaps even larger since our gaps (Tab.~\ref{tab:nf_gap}) are
slightly underestimated. On the other hand, the LDA+U method recovers
the correct band gaps in the low-temperature ordered phases already
with $U$ around 4.0~eV.\cite{suzuki2013} This difference originates in
the different ways the screening by oxygen ligands is accounted for in the
two methods. In LDA+DMFT, the oxygen states explicitly enter the
impurity solver as the bath and hence the $U$ used in the solver
does not include the screening by these states. The LDA+U method, on the
other hand, is a static limit of the LDA+DMFT where a simplified 
Hartree--Fock approximation \emph{in the atomic limit} plays the
role of the impurity solver. In this case, the screening by the ligand
states can enter the calculations only implicitly in the form of a
reduced $U$.

The so-called Hubbard~I
approximation\cite{hubbard1963,lichtenstein1998} is analogous to the
LDA+U method with respect to the screening of the on-site Coulomb
interaction. In the same time, it allows for a description of the
paramagnetic phase and hence can be straightforwardly compared to
our LDA+DMFT results. The impurity model for the Hubbard~I
approximation is given by Eq.~\eqref{eq:Himp0} with
$\mathbb{V}=0$ and $\mathbb{H}_{\rm
  bath}=0$. Figure~\ref{fig:UO2_dmft_hub1} shows that
the spectral density of UO${}_2$ from the LDA+Hubbard~I method
combined with a reduced $U=3.0$~eV is almost identical to the spectral
density from the LDA+DMFT 
calculations with $U=6.5$~eV and with all other inputs unchanged. It is
the screening by the ligand states that is responsible for the
difference in the Coulomb parameter. The same behavior is found also in
NpO${}_2$ and PuO${}_2$ (not shown). A related discussion of the
screening effects in transition metal oxides can be found in
Ref.~\onlinecite{haule2014}.

The close similarity between the LDA+Hubbard~I and LDA+DMFT
spectra of UO${}_2$ indicates that the main 5f peak near $-1$~eV
is a lower Hubbard band. It was argued in Ref.~\onlinecite{yin2011} that this
feature is a Zhang--Rice state resulting from a coupling
of the local moment in the 5f shell to a hole in the ligand
orbitals.\cite{zhang1988} But since the Hubbard~I approximation is
clearly able to describe this peak, it cannot originate in the Zhang--Rice
physics because the Hubbard~I selfenergy has no knowledge of the ligand
states.

\subsection{Core-level spectra}

\begin{figure*}
\resizebox{.9\linewidth}{!}{\includegraphics{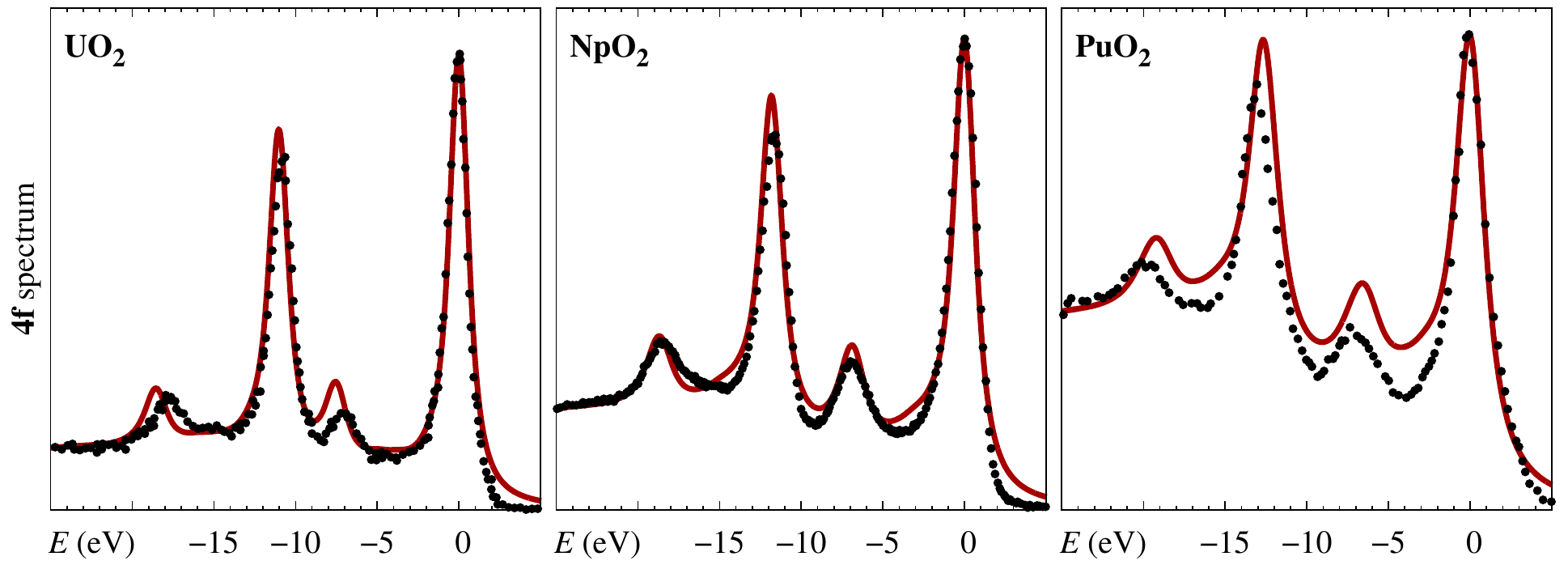}}
\caption{\label{fig:4fdos}(Color online) Calculated spectra of x-ray
  photoemission from the 4f core level (red lines) are compared with
  the experimental data (black dots) from
  Ref.~\onlinecite{baer1980} (UO${}_2$), Ref.~\onlinecite{teterin2014}
  (NpO${}_2$), and Ref.~\onlinecite{veal1977} (PuO${}_2$).
  The theoretical and experimental spectra are aligned at the
  4f$_{7/2}$ line that is placed at the zero of energy. The life-time
  broadening $\Gamma_{\rm c}$ was adjusted to match the width of the
  experimental 4f$_{7/2}$ line in each oxide separately. A background
  due to the secondarily scattered electrons was added to the theoretical
  curves as described in Ref.~\onlinecite{kotani1992}.}
\end{figure*}


The nominal electronic configuration of the actinide atoms in the
investigated dioxides is the $4+$ ion, which translates to the number
of 5f electrons $n_f$ equal two, three, and four in UO${}_2$,
NpO${}_2$, and PuO${}_2$, respectively. Our LDA+DMFT calculations
converge to solutions with more electrons in the 5f Wannier orbitals,
with $n_f$ being close to half-integer values, see
Tab.~\ref{tab:nf_gap}. A similarly increased filling of the 5f shell
was reported previously in LDA+U, hybrid DFT, and embedded-cluster
calculations.\cite{shick2014,roy2008,bagus2013} The enhancement of the
5f occupation indicates a large covalent mixing (hybridization)
between the actinide 5f states and oxygen 2p states. An evidence of
such mixing was found also in the experimental core-level spectra
where the hybridization with ligands is responsible for the appearance
of satellite peaks.\cite{kotani1992} An analogous half-integer
occupation of the valence f~shell is well established in CeO${}_2$ on
the basis of the core-level
spectroscopy.\cite{fujimori1983,wuilloud1984,kotani2012} Surprisingly,
the earlier LDA+DMFT study of the actinide dioxides reported an
integer occupation of the 5f shell in all of them.\cite{yin2011}

In order to test whether our estimates of $n_f$ are compatible with
the measured core-level spectra, we have calculated the photoemission
from the 4f states using the method outlined in
Sec.~\ref{sec:spectra_theo}. Since the theory is formulated for a
non-degenerate core state, the total 4f spectrum is approximated as a
sum of two independent components (4f${}_{5/2}$ and 4f${}_{7/2}$) that
are weighted with the statistical ratio $3:4$. The splitting between
the 4f${}_{5/2}$ and 4f${}_{7/2}$ levels is taken from the
all-electron LDA: 11.0~eV (UO${}_2$), 11.8~eV (NpO${}_2$), and~12.6 eV
(PuO${}_2$). We employ two empirical parameters in the calculations of
the core-level spectra: the core-valence Coulomb repulsion, $U_{\rm
cv}=6.0$~eV, and the scaling factor of the hybridization in the final
state, $q=0.85$. We keep these parameters the same for all three
oxides. Our results are plotted in Fig.~\ref{fig:4fdos} together with
the experimental spectra. The satellites are well reproduced which
indicates that the LDA+DMFT fillings $n_f$ are indeed reasonable. It
is possible to analyze individual contributions to each of the
spectral features,\cite{kotani1992,bagus2013} but we do not enter that
level of detail here.

\subsection{Crystal-field states}

\begin{table}
\caption{\label{tab:cf} Splitting of the lowest 5f multiplet by the
  cubic environment. The present calculations are compared to the data
  inferred from the inelastic neutron scattering experiments that were
  performed above the ordering temperature.}
\begin{ruledtabular}
\begin{tabular}{llclll}
&&ground state & \multicolumn{3}{c}{excited levels (meV)}\\[-.15em]
&&multiplicity & \multicolumn{3}{c}{(multiplicity in brackets)}\\
\hline
UO${}_2$  & theory & 3 & 161 (1) & 164 (3) & 184 (2) \\
 & Ref.~\onlinecite{nakotte2010} && \multicolumn{2}{c}{150} & 170 \\
NpO${}_2$ & theory & 4 & \phantom{0}76 (4) & 302 (2) & \\
 & Ref.~\onlinecite{amoretti1992} && \phantom{0}55 & \phantom{00}-- & \\
PuO${}_2$ & theory & 1 & 125 (3) & 226 (3) & 319 (2) \\
 & Ref.~\onlinecite{kern1999} && 123 & \phantom{00}-- & \phantom{00}-- \\
\end{tabular}
\end{ruledtabular}
\end{table}

Finally, we discuss the role of the cubic environment around the
actinide atoms. Its most important implications are the complex
multipolar order in the low-temperature phases of UO${}_2$ and
NpO${}_2$,\cite{santini2009} and the absence of magnetism in
PuO${}_2$.\cite{raphael1968} The ordered states are outside the scope
of the present study. The case of PuO${}_2$ was discussed in our
earlier paper -- the cubic environment splits the 5f shell such that
the ground state is non-degenerate, and the temperature dependence of
the magnetic susceptibility, which appears due to the thermal
population of excited states, is reduced due to a cancellation of the
spin and orbital contributions to the susceptibility.\cite{shick2014}

Experimentally, the crystal field can be probed by inelastic neutron
scattering (INS) that reflects the local electronic structure at the
actinide atoms. Typically, a few lowest excited states can be
detected. In our theoretical description, these states correspond to
the bottom of the many-body spectrum of the converged DMFT impurity
model. We compare the computed crystal-field splitting of the lowest
5f multiplet with the findings of the INS experiments in
Tab.~\ref{tab:cf}. In the paramagnetic state of UO${}_2$, the
experiments detect only two excitations.\cite{nakotte2010} We obtain
three but the first two of them are nearly degenerate and hence they
appear as one peak in the experimental spectrum. Indeed, this peak is
found to split into two at the transition into the low-temperature
ordered phase.\cite{nakotte2010} In NpO${}_2$ and PuO${}_2$, only the
first crystal-field excitation is observed in the INS
experiments.\cite{amoretti1992,kern1999} The theoretical prediction is
in excellent agreement in PuO${}_2$ but it is less accurate in the
case of NpO${}_2$. A limited accuracy of the present theory is
expected due to the neglect of the non-spherical components of the
double counting correction as discussed near the end of
Sec.~\ref{sec:DMFT}. The good performance of the theory in PuO${}_2$
is likely to originate in the highly symmetric 5f charge density in
the LDA solution.\cite{suzuki2013}

Our calculations yield the same ground-state multiplicity as
an alternative first-principles calculation based on the LDA+U
method,\cite{zhou2012} but the ordering of some of the excited states in
UO${}_2$ and PuO${}_2$ is different. A direct comparison of the
crystal-field parameters between Ref.~\onlinecite{zhou2012} and our theory
is not possible since we have non-spherical terms not only in the the
crystal-field potential inside $\mathbb{H}_{\rm loc}$ but also in the
hybridization with the ligand states, each contributing approximately
half of the splitting due to cubic environment.\cite{kolorenc2015}

\section{Conclusions}

We have demonstrated that an implementation of the LDA+DMFT method
where the selfenergy is obtained by the exact diagonalization of a
finite impurity model provides an accurate description of the
electronic structure of the early actinide dioxides in the
paramagnetic phase. The band-gap opening is not linked to any
long-range order, and the main features of the valence-band as well as
the core-level photoemission spectra are well reproduced. The method
allows for a quantitative first-principles analysis of the covalency
between the actinide 5f and oxygen 2p states that is found to induce a
nearly half-integer filling of the actinide 5f shell.

\begin{acknowledgments}
We acknowledge financial support from the Czech-German collaboration
program (GACR 15-05872J). Access to computing and storage facilities
owned by parties and projects contributing to the National Grid
Infrastructure MetaCentrum, provided under the programme ``Projects of
Large Infrastructure for Research, Development, and Innovations''
(LM2010005), is appreciated.
\end{acknowledgments}

\appendix

\section{Construction of the impurity model}
\label{app:bath_discretization}

In this appendix we discuss how the parameters of the finite
non-interacting impurity model from Eq.~\eqref{eq:Himp0} are found so
that the model matches the effective medium (the bath) as closely
as possible. The hamiltonian has the form a block matrix
\begin{equation}
\mathbb{H}_{\rm imp}^{(0)}=
\begin{pmatrix}
\mathbb{H}_{\rm loc} & \mathbb{V} \\
\mathbb{V}^{\dagger} & \mathbb{H}_{\rm bath}
\end{pmatrix},
\end{equation}
where all blocks are $14\times 14$ square matrices. 
They are determined by comparing the large $z$ asymptotics
of the local block of the Green's function,\cite{Note1}
\begin{equation}
\label{eq:Gloc}
\mathbb{G}_{\rm loc}^{(0)}(z)=\Bigl[z -\mathbb{H}_{\rm loc}
 -\mathbb{V}\bigl(z -\mathbb{H}_{\rm bath}\bigr)^{-1}\mathbb{V}^{\dagger}
\Bigr]^{-1},
\end{equation}
to the asymptotics of
the so-called bath Green's function $\mathbb{G}_0(z)$,
which is the local Green's function corresponding to the
effective medium $\bigl(\hat H_{\rm TB}+\hat\selfen\bigr)$. The
procedure follows the steps outlined in Ref.~\onlinecite{kolorenc2012a} with the
notable difference that the local hamiltonian~$\mathbb{H}_{\rm loc}$ now
contains a strong spin-orbital coupling that does not commute with the
cubic hybridization function $\mathbb{V}\bigl(z
-\mathbb{H}_{\rm bath}\bigr)^{-1}\mathbb{V}^{\dagger}$.
Therefore, the problem cannot be simplified to diagonal matrices.

\begin{figure}
\resizebox{0.65\linewidth}{!}{\includegraphics{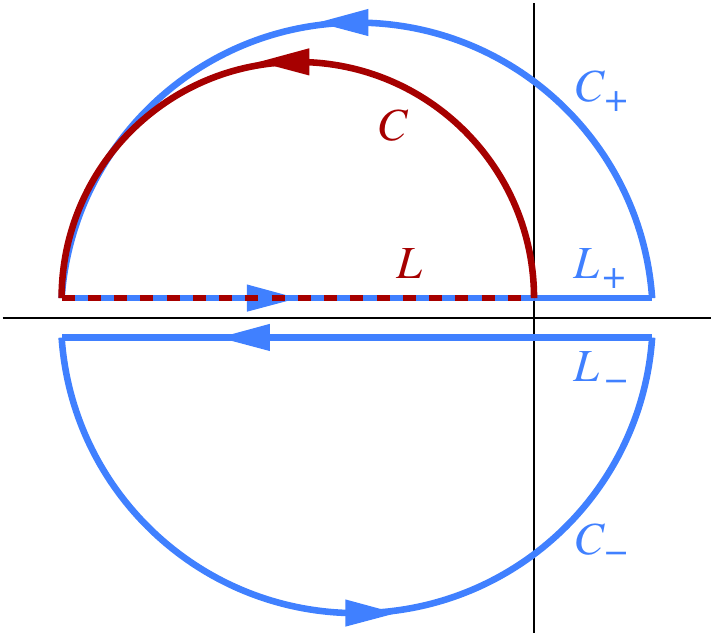}}
\caption{\label{fig:moment_contours}(Color online) Contours in the
  complex plane used for integration of the moments~$\mathbb{M}_n$
  (shown in blue). Line segments are denoted as $L_{\pm}$, half
  circles as $C_{\pm}$. The smaller contour shown in red is employed
  to calculate the band occupations during the DMFT cycle (the
  vertical line indicates the Fermi level).}
\end{figure}

The asymptotic expansion of $\mathbb{G}_{\rm loc}^{(0)}(z)$ can be
found by a repeated application of the identity 
\begin{equation}
\bigl(\mathbb{A}-\mathbb{B}\bigr)^{-1}=\mathbb{A}^{-1}
  +\mathbb{A}^{-1}\mathbb{B}\bigl(\mathbb{A}-\mathbb{B}\bigr)^{-1},
\end{equation}
which yields
\begin{multline}
\mathbb{G}_{\rm loc}^{(0)}(z)\approx\frac1z
+ \frac{\mathbb{H}_{\rm loc}}{z^2}
+\frac1{z^3}\,
\bigl(\mathbb{H}_{\rm loc}^2+\mathbb{V}\mathbb{V}^{\dagger}\bigr)
+\frac1{z^4}\bigl(\mathbb{H}_{\rm loc}^3\\[.2em]
+\mathbb{H}_{\rm loc}\mathbb{V}\mathbb{V}^{\dagger}
+\mathbb{V}\mathbb{V}^{\dagger}\mathbb{H}_{\rm loc}
+\mathbb{V}\,\mathbb{H}_{\rm bath}\mathbb{V}^{\dagger}\bigr)\,,
\label{eq:GFimp_asympt}
\end{multline}
where we kept only contributions up to $1/z^4$.
From the other side, the bath Green's function in the spectral
representation reads as 
\begin{equation}
\mathbb{G}_0(z)
=\int\frac{\mathbb{A}_0(\epsilon)}{z-\epsilon}\,\rmd\epsilon\,,
\label{eq:GF0spectral}
\end{equation}
where we introduced the spectral density of the bath,
\begin{equation}
\mathbb{A}_0(\epsilon)
=\frac{\mathbb{G}_0(\epsilon-\rmi0)-\mathbb{G}_0(\epsilon+\rmi0)}{2\pi\rmi}\,.
\end{equation}
The asymptotic expansion of the bath
Green's function is obtained by expanding the denominator in
Eq.~\eqref{eq:GF0spectral},
\begin{equation}
\mathbb{G}_0(z)=\sum_{n=0}^{\infty}\frac{\mathbb{M}_{n}}{z^{n+1}}\,,
\quad
\mathbb{M}_{n}=\int\epsilon^{n} \mathbb{A}_0(\epsilon)\,\rmd\epsilon\,,
\label{eq:GF0_asympt}
\end{equation}
where $\mathbb{M}_n$ are moments of the spectral density. The
spectral density $\mathbb{A}_0(\epsilon)$ is a hermitian matrix and
hence its moments are hermitian matrices as well. The
moments can be written in terms of contour integrals in 
the complex plane. Using the path segments depicted in
Fig.~\ref{fig:moment_contours} we have
\begin{multline}
\mathbb{M}_n
=\frac1{2\pi\rmi}\biggl[\int_{-L_-}-\int_{L_+}\biggr]\,
  z^n\mathbb{G}_0(z)\, \rmd z\\
=\frac1{2\pi\rmi}\biggl[\int_{C_-}+\int_{C_+}\biggr]\,
  z^n\mathbb{G}_0(z)\, \rmd z\,,
\end{multline}
that is, an integral over a circle $C_-\cap C_+$ that encloses the
entire support of $\mathbb{A}_0(\epsilon)$. In the calculations
reported in this paper we encircle the real-axis segment from $-20$~eV to
$10$~eV. 

When the bath contains the same number of orbitals as the impurity
shell, the blocks $\mathbb{H}_{\rm loc}$, $\mathbb{H}_{\rm bath}$, and
$\mathbb{V}$ are all square matrices of the same size. It is then
straightforward to construct these matrices by comparing the
asymptotic expansions, Eqs.~\eqref{eq:GFimp_asympt}
and~\eqref{eq:GF0_asympt}, term by term. The $1/z^2$ terms give the
local hamiltonian,
\begin{subequations}
\begin{equation}
\mathbb{H}_{\rm loc}=\mathbb{M}_1\,.
\end{equation}
The $1/z^3$ terms lead to the equation $\mathbb{V}\mathbb{V}^{\dagger}
=\mathbb{M}_2-\mathbb{M}_1^2$ which has a solution 
\begin{equation}
\mathbb{V}=\mathbb{V}^{\dagger}=\sqrt{\mathbb{M}_2-\mathbb{M}_1^2}
\end{equation}
as long as all eigenvalues of $\mathbb{M}_2-\mathbb{M}_1^2$ are
non-negative. If some of them were negative, the bath Green's function
would not be representable by an impurity model, since the product
$\mathbb{V}\mathbb{V}^{\dagger}$ is always 
positive as follows from $\langle\psi|\mathbb{V}
\mathbb{V}^{\dagger}|\psi\rangle =\langle\mathbb{V}^{\dagger}\psi|
\mathbb{V}^{\dagger}\psi\rangle\geq 0$ for any
$|\psi\rangle$. Finally, the $1/z^4$ terms coincide if we set
\begin{equation}
\mathbb{H}_{\rm bath}=\mathbb{V}^{-1}\bigl[\mathbb{M}_3+\mathbb{M}_1^3
-\mathbb{M}_1\mathbb{M}_2-\mathbb{M}_2\mathbb{M}_1\bigr]
\bigl(\mathbb{V}^{\dagger}\bigr)^{-1}\,.
\end{equation}
\label{eq:Himp14x14}%
\end{subequations}
Equations~\eqref{eq:Himp14x14} define all blocks of $\mathbb{H}_{\rm
  imp}^{(0)}$ as hermitian matrices. Alternatively, we could take
advantage of the freedom to arbitrarily choose the basis in the bath
segment and diagonalize $\mathbb{H}_{\rm bath}$ instead. The
transformation reads as
\begin{equation}
\mathbb{H}_{\rm bath}\to
  \mathbb{C}^{-1}\mathbb{H}_{\rm bath}\mathbb{C}\,,\quad
\mathbb{V}\to \mathbb{V}\mathbb{C}\,,
\end{equation}
where $\mathbb{C}$ is the appropriate unitary matrix and the new
$\mathbb{V}$ is no longer hermitian.


\section{Lanczos impurity solver in a reduced many-body basis}
\label{app:solver}

The selfenergy $\hat\selfen(z)$ is computed from the Green's function of
the interacting impurity model $\hat H_{\rm imp}$ as
\begin{equation}
\label{eq:selfen}
\hat\selfen(z)=z-\hat H_{\rm imp}^{(0)}
-\hat G^{-1}(z)\,.
\end{equation}
Since the Coulomb vertex $\hat U$ defined in Eq.~\eqref{eq:vertex}
only acts among the impurity orbitals, the selfenergy is non-zero only
in the local block $\bbselfen_{\rm loc}(z)$. Consequently, it
is only the local block of the Green's function that needs to be
explicitly evaluated,
\begin{equation}
\label{eq:selfenLoc}
\bbselfen_{\rm loc}(z)=z-\mathbb{H}_{\rm loc}
-\mathbb{V}\bigl(z-\mathbb{H}_{\rm
  bath}\bigr)^{-1}\mathbb{V}^{\dagger}-\mathbb{G}_{\rm loc}^{-1}(z)\,.
\end{equation}
Simplification to a diagonal representation is not possible
because $\mathbb{H}_{\rm loc}$ and $\mathbb{V}\bigl(z-\mathbb{H}_{\rm
  bath}\bigr)^{-1}\mathbb{V}^{\dagger}$ do not commute.

The Green's function reads as
\begin{subequations}
\label{eq:GFgrandcanonical}
\begin{equation}
\mathbb{G}_{\rm loc}(z)=
\sum_N\sum_{\alpha}w_{N\alpha}\bigl[
\mathbb{G}_{N\alpha}^{>}(z)
+\mathbb{G}_{N\alpha}^{<}(z)\bigr]\,,
\end{equation}
where the two components are
\begin{align}
\bigl[\mathbb{G}_{N\alpha}^{>}(z)\bigr]_{ij}&=
 \langle N\alpha|\hat f_{i}
\frac1{z+E_{N\alpha}-\hat H_{\rm imp}}
 \hat f^{\dagger}_{j}|N\alpha\rangle\,, \\
\bigl[\mathbb{G}_{N\alpha}^{<}(z)\bigr]_{ij}&=
 \langle N\alpha|\hat f^{\dagger}_{i}
\frac1{z-E_{N\alpha}+\hat H_{\rm imp}}
 \hat f_{j}|N\alpha\rangle\,.
\end{align}
\end{subequations}
The sums represent the grandcanonical average, they run over the
filling of the impurity model $N$ and over the many-body spectrum
$\alpha$, $\hat H_{\rm
  imp}|N\alpha\rangle=E_{N\alpha}|N\alpha\rangle$. The spectrum is
calculated independently for each filling because a separate
Hilbert space $\mathcal{H}_N$ is associated with each~$N$. The
grandcanonical weights have the form
\begin{equation}
w_{N\alpha}=
\rme^{-\beta E_{N\alpha}}\Big/
\sum_N\sum_{\alpha}\rme^{-\beta E_{N\alpha}}\,,
\end{equation}
where we set the chemical potential to zero without any loss of
generality. At low temperatures $T$ (large $\beta=1/T$), only a few
weights $w_{N\alpha}$ have an appreciable magnitude and hence both
sums in Eqs.~\eqref{eq:GFgrandcanonical} can be truncated for
an increased computational efficiency.

Practically, the bottom of the spectrum including all degeneracies is
found for each relevant $N$ using the implicitly restarted Lanczos
method implemented in the \code{ARPACK} software package.\cite{arpack}
The matrix elements of the Green's function,
Eqs.~\eqref{eq:GFgrandcanonical}, are evaluated with the aid of the
band Lanczos method.\cite{ruhe1979,meyer1989} The band variant is a
convenient way to access diagonal as well as off-diagonal matrix
elements of $\mathbb{G}_{\rm loc}(z)$.

The Lanczos method for
evaluation of matrix elements of a resolvent, $\bigl(z-\hat
H_{\rm imp}\bigr)^{-1}$, is essentially an expansion in powers of
$\hat H_{\rm imp}$. The length of the expansion is given by the size of
the Krylov space.\cite{parlett,meyer1989} The farther is the
complex energy $z$ from the spectrum of $H_{\rm imp}$, the shorter the
expansion needs to be. During the DMFT iterations, we evaluate
the selfenergy and the Green's function along the three
semicircular contours indicated in Fig.~\ref{fig:moment_contours},
$C_{\pm}$ to determine the non-interacting impurity model and $C$ to
check the number of electrons below the Fermi level. In an insulator,
the closest any of these contours gets to the spectrum of $H_{\rm
  imp}$ is half of the band gap (contour~$C$) and hence one can get away
with a relatively small Krylov space.

Despite the high efficiency of the Lanczos method, the calculations as
outlined so far would be very demanding, especially on memory, because
the many-body Hilbert spaces $\mathcal{H}_N$ are very large.  As it
turns out, the Hilbert spaces can be substantially reduced without
compromising the accuracy of the calculations. To this end, we take
inspiration in the method developed for Ce compounds by Gunnarsson and
Sch\"onhammer,\cite{gunnarsson1983} which can be viewed as an
expansion in the hybridization parameters~$\mathbb{V}$ around the
atomic limit.


Let us take the ground state of the impurity model in UO${}_2$ as an
example. The bath orbitals represent the oxygen 2p states and hence
they are located several eV below the Fermi level (recall the band
structure from Fig.~\ref{fig:UO2_LDA_bands}). If there were no
hybridization between the bath and the f~shell, the bath would be
completely occupied with 14 electrons and there would be 2 electrons
in the f~shell. Hence, the ground state would be a linear combination
of basis states $|f^2\text{\underline{$b$}}^0\rangle$ where
\underline{$b$} indicates a hole in the bath orbitals. If only these
states are considered, the bath degrees of freedom are completely
frozen and the problem is reduced to the atomic limit, the so-called
Hubbard-I approximation. When the hybridization is present, the states
$|f^2\text{\underline{$b$}}^0\rangle$ mix with states
$|f^{2+m}\text{\underline{$b$}}^m\rangle$ where $m$ electrons hopped
from the bath to the f shell. As the hybridization increases, basis
states with larger and larger $m$ become relevant and need to be taken
into account. The states $|f^{2+m}\text{\underline{$b$}}^m\rangle$ can
be obtained from the zeroth-order approximation
$|f^2\text{\underline{$b$}}^0\rangle$ by a repeated application of the
hybridization part of the impurity hamiltonian,
\begin{equation}
\sum_{iJ}\Bigl(
\bigl[\mathbb{V}\bigr]_{iJ}\hat f_i^{\dagger}\hat b_J
+\bigl[\mathbb{V}^{\dagger}\bigr]_{Ji}\hat b_J^{\dagger}\hat f_i
\Bigr)\,,
\end{equation}
hence the connection with the perturbation expansion in
$\mathbb{V}$. Following the outlined logic, we define truncated Hilbert spaces
\begin{equation}
\label{eq:truncHilbert}
\mathcal{H}_N^{(M)}=\bigl\{
|f^{N-14+m}\,\text{\underline{$b$}}^m\rangle,\, 0\leq m\leq M
\bigr\}
\end{equation}
and perform the many-body calculations only in these reduced
spaces. For the actinide dioxides, we have explored
$M$ up to 4 and have found that $M$ equal 3 is already enough
to reach essentially converged results for all quantities discussed in
this paper. The Hilbert-space dimensions can thus be reduced from as
high as $37 \times 10^6$ in the case of the full $\mathcal{H}_N$ down
to about $1.5 \times 10^6$ in the case of $\mathcal{H}_N^{(3)}$.


The same truncation of the Hilbert space can be considered in a
general impurity model with $N_b^<$ bath states below the Fermi level
and $N_b^>$ bath states above the Fermi level. The zeroth-order Hilbert
space corresponding to the atomic limit then is
$|f^{N-N_b^<}b^0\text{\underline{$b$}}^0\rangle$ where $b$ indicates
electrons in the bath orbitals above the Fermi level. The general
truncated Hilbert spaces then read as
\begin{equation}
\label{eq:truncHilbertGeneral}
\mathcal{H}_N^{(M)}=\bigl\{
|f^{N-N_b^<-n+m}\,b^n\,\text{\underline{$b$}}^m\rangle,\,
0\leq m+n\leq M\bigr\}\,.
\end{equation}
We used them in the past to study metallic f-electron
compounds.\cite{gorelov2010,shick2013} The approximation is
variational, the truncated Hilbert space $\mathcal{H}_N^{(M)}$ turns
into the full space $\mathcal{H}_N$ for large enough $M$. Therefore,
although the reduction of the many-body basis was formulated as an
expansion around the atomic limit, it can prove useful also in cases
where the atomic limit itself is a poor approximation.

The reduced basis comes with one disadvantage: the selfenergy is
non-zero also outside the local block $\bbselfen_{\rm loc}$ and hence
Eq.~\eqref{eq:selfenLoc} does not hold. What is worse,
Eq.~\eqref{eq:selfenLoc} cannot be used even as an approximation
because it leads to a non-causal selfenergy (alternating sign of the
imaginary part of $\Sigma$). Consequently, the selfenergy has to be
calculated directly from Eq.~\eqref{eq:selfen} that involves a larger
Green's function matrix ($28\times 28$ in our case). This, in turn,
necessitates a twice wider band in the band Lanczos method when
compared to the calculation in the full Hilbert space
$\mathcal{H}_N$. Still, the associated increase of computational
complexity is significantly outweighed by the savings offered by the
smaller Hilbert space.

The appearance of non-vanishing selfenergy outside the local block can
be understood using the following argument. The cutoff in
Eq.~\eqref{eq:truncHilbert} can be implemented ``dynamically'' by
introducing an artificial multi-body repulsion between holes in the bath
orbitals. The extra term in the hamiltonian can be schematically
written as
\begin{equation}
\Delta\hat H\sim U_P
\underbrace{\hat b\dots\hat b}_{M+1}\:
\underbrace{\hat b^{\dagger}\dots\hat b^{\dagger}}_{M+1}\,,
\end{equation}
which adds a penalty $U_P$ to the energy of states with more than $M$
holes in the bath. The strict cutoff is achieved in the limit
$U_P\to\infty$. The interaction $\Delta\hat H$ apparently induces a
selfenergy to the bath block, $\bbselfen_{\rm bath}\not=0$.

Recently, a conceptually similar technique for reduction of the many-body
basis was introduced in Ref.~\onlinecite{lu2014}. It is substantially more
sophisticated than the method we describe here in that it does not
require any a priori chosen cutoff $M$. Nevertheless, it was
demonstrated to work only in a single-orbital impurity model so far, and it
remains to be seen how it performs in realistic DMFT calculations
with multiorbital impurity models.


\bibliography{dft,dmft,xpsDMFT,lanczos,aim,codes,AnO,ReO,pu,tmo}

\end{document}